\documentclass[11pt]{article}
\input epsf.sty
\usepackage{amsmath}
\usepackage{cite}
\usepackage{amssymb}
\usepackage{mathtools}
\usepackage{mathrsfs}
\usepackage{bm}
\usepackage{graphicx}
\usepackage{mathrsfs}
\usepackage{amsmath}
\usepackage{mathrsfs}
\usepackage{amssymb}
\usepackage{color}
\usepackage{hyperref}
\usepackage{psfrag}

\newcommand{\bq}{\begin{eqnarray}}
\newcommand{\eq}{\end{eqnarray}}

\begin{document}
\topmargin 0mm
\oddsidemargin 0mm
\begin{titlepage}
\begin{center}
\hfill USTC-ICTS-13-14\\
\hfill October 2013
\vspace{2.5cm}
%\begin{center}

{\Large \bf Higgs decays to $\gamma$ and invisible particles in the standard model}

\vspace{10mm}

{Yi Sun$^{\dagger}$ and Dao-Neng Gao$^{\ddagger}$ }
\vspace{4mm}\\{\it\small
Interdisciplinary Center for Theoretical Study,
University of Science and Technology of China, Hefei, Anhui 230026 China}\\
\vspace{4mm}
\end{center}

\vspace{10mm}
\begin{abstract}
\noindent
Using the Higgs boson mass $m_h=125$ GeV, the radiative Higgs decays $h\rightarrow\gamma \nu_l\bar\nu_l$ with
$\nu_l = \nu_e,\,\nu_\mu$ and $\nu_\tau$ are analyzed in the standard
model. Our calculation shows that the inclusive width of these processes, i.e., the sum of $\Gamma(h\to\gamma\nu_l\bar\nu_l)$ for ${\nu_l=\nu_e,\nu_\mu,\nu_\tau}$, is $1.41$ keV, which is about $15\%$ of $\Gamma(h\to\gamma\gamma)$.
Therefore, the observation of these channels in the future precise experiments may provide us some useful information on the Higgs physics both in the standard model and in its possible extensions.
%Thus these channels might be interesting in the future precise experiments both to test our understanding %of Higgs physics in the standard model and to probe the novel Higgs dynamics in new physics scenarios.
\end{abstract}
\vfill
\noindent
$^{\dagger}$ E-mail:~sunyi@mail.ustc.edu.cn\\
\noindent
$^{\ddagger}$ E-mail:~gaodn@ustc.edu.cn

\end{titlepage}
%\newpage
The discovery of the Higgs-like particle at around 125 GeV, thanks to the hard work of ATLAS \cite{atlas} and CMS \cite{cms} Collaborations at the Large Hadron Collider (LHC), is a big triumph of the high energy physics community.
Since elementary particles in the standard model (SM) become massive via the Higgs mechanism \cite{higgs}, the Higgs sector plays a key role in our understanding of the nature of the world.
Thus it is very important to identify the new resonance with the elementary Higgs boson in the SM.
ATLAS \cite{atlas} and CMS \cite{cms} have observed several decay channels of this Higgs-like particle, including $\gamma\gamma$, $ZZ^*$, $WW^*$, $bb$ and $\tau\tau$ channels, and have found that its properties are consistent with the SM Higgs boson.

With the increasing of the experimental data, besides the above dominant decay channels, some rare decay modes will also be interesting.
The radiative decays $h\to\gamma l^+l^-$ have been analyzed both theoretically \cite{ABDR96, LZQ98, AR00, onlyonechannel, CQZ12, DR13, KK13, sun13,Passarino, DICUS13} and experimentally\cite{atlasreport, cmsreport}. In the present paper, we will study another
rare decay channels  $h\to\gamma\nu_l\bar\nu_l$ with $\nu_l=\nu_e,\,\nu_\mu$ and $\nu_\tau$. Obviously, only a photon, or more accurately, a photon and missing energy\cite{missingenergy}, can be observed in these decays experimentally.
On the other hand, in the new physics, the decay channel $h\to\gamma+invisible$ is more complex than that in the SM, where the invisible particles could be, besides neutrinos, other new particles which are absent in the SM. Model independently, in Ref. \cite{KC12}, the Higgs decay mode involving a photon together with one or two invisible particles has been investigated using effective interactions. In some specific models, the mode $h\to\gamma Z_d$, where $Z_d$ is a light vector boson associated to a ``dark sector" U(1) gauge group, has been analyzed in \cite{DLLM13};
in the next to minimal supersymmetric standard model, the process $h\to\gamma\chi_1\chi_1$, where $\chi_1$ is the lightest supersymmetry particle and invisible in the experiments, may be interesting in some parameter space \cite{tliu13}.
Therefore, in order to analyze these exotic decays in the new physics beyond the SM, we should first evaluate their contributions in the SM. Only after we fully understand their SM background, the future precise experimental study of the $h\to\gamma\nu_l\bar\nu_l$ decays might provide us some useful information on the Higgs sector in new physics scenarios.

\begin{figure}
  % Requires \usepackage{graphicx}
 \begin{center}
  \includegraphics[width=16cm,height=3cm]{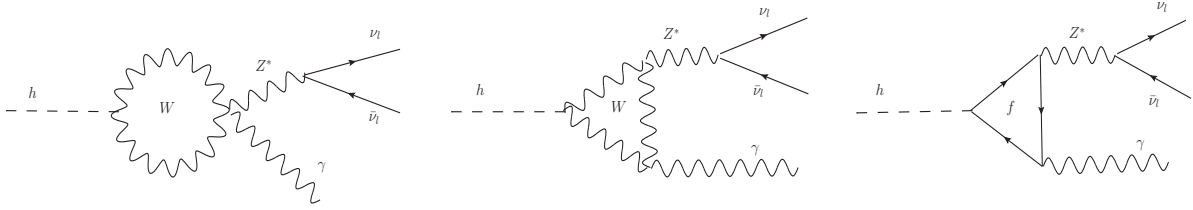}\end{center}
  \caption{Three-point diagrams for $h\rightarrow\gamma \nu_l\bar\nu_l$.}\label{three}
\end{figure}

\begin{figure}
\begin{center}
  \includegraphics[width=16cm,height=5cm]{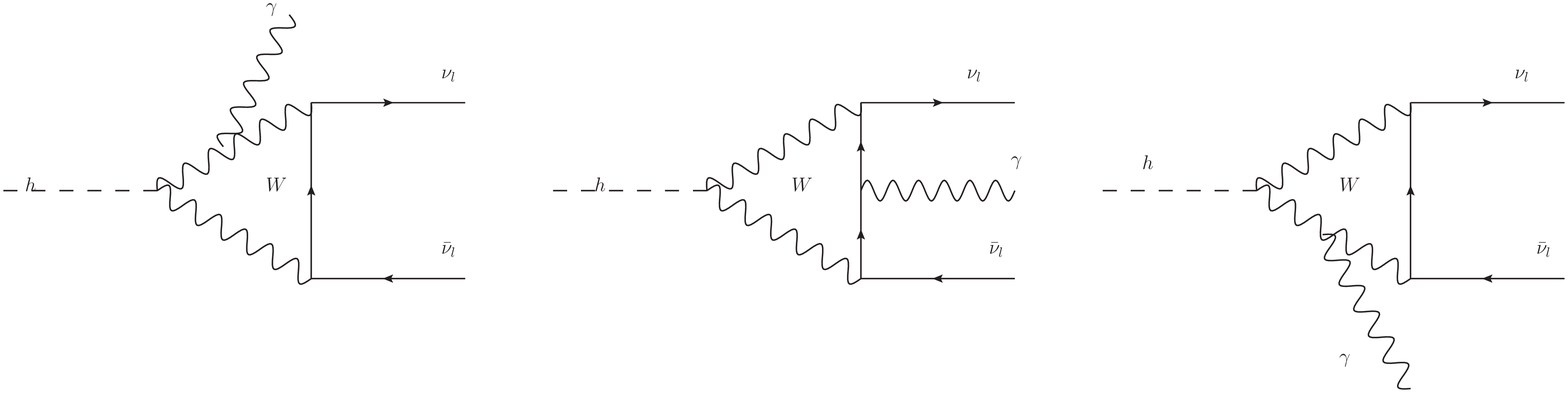}
  \end{center}
  \caption{Four-point diagrams for $h\rightarrow\gamma \nu_l\bar\nu_l$.}\label{four}
\end{figure}

In the SM, the tree-level contribution of the processes $h\to\gamma\nu_l\bar\nu_l$ is forbidden and the lowest order contribution is given by the one-loop diagrams.
The typical one-loop Feynman diagrams for these processes have been shown
in Figures \ref{three} and \ref{four}, respectively, which are of two basic
types: (i) the $Z^*$ pole three-point diagrams via $h\to\gamma Z^*\to\gamma\nu_l\bar{\nu_l}$ (Figure \ref{three}); (ii) four-point box diagrams involving $W$ gauge boson and the charged lepton $l$ inside the loop (Figure \ref{four}), where the photon is emitted from the $W$ or $l$ internal lines.

The amplitude of $h\to\gamma\nu_l\bar\nu_l$ at the one-loop level can be expressed as
\begin{eqnarray}
  \mathcal{M}&=&\mathcal{M}_{tri}+\mathcal{M}_{box},
\end{eqnarray}
where $\mathcal{M}_{tri}$ and $\mathcal{M}_{box}$ denote the amplitudes of the three-point and four-point diagrams, respectively.
\begin{eqnarray}
\mathcal{M}_{tri}&=&\varepsilon^{\nu\ast}(p)C_1
\left(p_{\mu }q_{\nu }-g_{\mu \nu }p\cdot q \right)\bar {u}(k_2) \gamma^{\mu}P_Lv(k_1)\label{triangle},\\
\mathcal{M}_{box}&=&\varepsilon^{\nu\ast}(p)\bar {u}(k_2)\left[\left(C_2 k_1^{\nu}+ C_3 k_2^\nu\right) p\!\!\!/P_L-
\left( C_2 \,k_1\cdot p+ C_3\, k_2\cdot p\right) \gamma^{\nu}P_L\right]v(k_1),\label{box}
\end{eqnarray}
with $P_L=%\frac{1-\gamma^5}{2}
(1-\gamma^5)/2$.
One can check that the amplitudes of the three-point and four-point diagrams are separately gauge invariant.
The expressions of $C_i$'s in the amplitudes are
\begin{eqnarray}
C_1&=&P_Z\frac{\alpha_e^2}{\sqrt{2}m_W\sin^3\theta_W}\left[-\frac{\cos\theta_W}{2m_W^2}I_1-\frac{2N_c^fQ_fm_f^2}{\left( m_h^2 - q^2\right)^2\cos\theta_W}\left( T_f-2Q_f\sin^2\theta_W\right)I_2\right],\\
C_2&=&\frac{-\alpha_e^2}{2m_W\sin^3\theta_W}I_3,\\
C_3&=&\frac{-\alpha_e^2}{2m_W\sin^3\theta_W}I_4.
\end{eqnarray}
Here $\alpha_e$ is the fine-structure constant and $\theta_W$ is the electroweak mixing angle.
$m_f$ is the mass, $N_c^f$ is the color multiplicity, $Q_f$, in unit of e, is the charge, $T_f$ is the third component of weak isospin of the fermion $f$ inside the fermion loop in Figure \ref{three}.
$k_1, k_2$ and $p$ represent the momentum of $\nu_l, \bar\nu_l$ and $\gamma$ in the final states, respectively. We denote $q$ as the momentum of the virtual particle $Z^*$ in Figure \ref{three}, $q^2=(k_1+k_2)^2$ is neutrino pair mass squared, and
$P_Z$ is from the propagator of the virtual $Z^*$ gauge boson, which reads
\begin{eqnarray}\label{pz}
P_Z=\frac{1}{q^2-m_Z^2+i m_Z \Gamma_Z}.
\end{eqnarray}
The notations $I_i$'s are given by
\begin{eqnarray}\label{ampofvarious}
I_1&=&-8m_W^2(4m_W^2-q^2)C^0_0-4(12m_W^4+4m_W^2q^2-q^2m_h^2)(C^0_1
+C_{11}^0+C_{12}^0),\\
I_2&=&-2q^2B_0(m_h^2,m_f^2,m_f^2)+2q^2B_0(q^2,m_f^2,m_f^2)\nonumber\\&&
+(m_h^2-q^2)
(-2+(m_h^2-q^2-4m_f^2)C_0(0,q^2,m_h^2,m_f^2,m_f^2,m_f^2)),\\
%\frac{\alpha _e^{2}\cos\theta_W}{\sqrt{2}m_W\sin\theta_W}[-\cot \theta_W A_W\left(\tau_W,\lambda_W\right)
%-2 N_c Q_f\frac{ T_f-2Q_f \sin^2\theta_W}{\sin\theta_W\cos\theta_W}A_{f}\left(\tau _f,\lambda _f\right)]\\
I_3&=&-4C^4_0+2C^2_1-2C^2_2+4(1+a+b+f)C^4_{12}
+l^2(C^3_2-C^1_{12}+C^3_{12}+C^3_{22})+2m_W^2\nonumber\\
&&\left[3D^2_1+2D^2_2
-D^2_3+D^3_1+D^3_3-l^2(D^1_0+D^2_0+D^3_0)
-(-2+l^2(b+f))D^1_{2}-(2\right.\nonumber\\&&\left.+l^2+al^2)D^1_3+
(2+l^2(1+a+b+f)(D^3_{23}-D^1_{23}
-D^2_{23}-D^1_{33}-D^2_{33}-D^2_{2}))\right],\\
I_4&=& -2(C^2_0+C^2_1+2C^2_2+2C^4_0)+4(1+a+b+f)C^4_{12}+l^2(C^5_1+C^5_{12}+C^5_{22}+\nonumber\\&&C^6_2+
C^6_{12}+C^6_{22})-2m_W^2\left[l^2
(D^1_0+D^2_0+D^3_0)+2(D^1_1+D^2_1+D^3_1+D^3_2)\right.
\nonumber\\&&\left.+(2+l^2(1+a+b+f))(D^1_{12}+D^1_{13}
+D^2_{13}-D^3_{2}-D^3_{12}-D^3_{22})\right].
\end{eqnarray}
with some dimensionless parameters: $a=k_1\cdot p/m_W$, $b=k_2\cdot p/m_W$, $f=k_1\cdot k_2/m_W$, and $l=m_l/m_W$, where $m_l$ is the mass of the charged lepton in the four-point diagrams\footnote{One should take care of the difference between $m_f$, which is the mass of the fermion inside the fermion-loop of the three-point diagrams, and $m_l$, which is the mass of the charged lepton in the four-point diagrams.}. The notations $C_i^j$'s and $D_i^j$'s read
\begin{eqnarray}
C_i^0&=&C_i(0,q^2,m_h^2,m_W^2,m_W^2,m_W^2),\nonumber\\
C_i^1&=&C_i(0,2k_1\cdot p,0,m_l^2,m_l^2,m_W^2),\,\,\,\,\,\,C_i^2=C_i(0,m_h^2,2k_2\cdot p,m_l^2,m_W^2,m_W^2),\nonumber\\
C_i^3&=&C_i(0,0,2k_1\cdot p,m_w^2,m_W^2,m_l^2),\,\,\,\,\,\,
C_i^4=C_i(0,m_h^2,2k_1\cdot k_2,0,m_W^2,m_W^2,m_W^2),\nonumber\\
C_i^5&=&C_i(2k_2\cdot p,0,0,m_l^2,m_W^2,m_W^2),\,\,\,\,\,\,
C_i^6=C_i(0,0,2k_2\cdot p,m_W^2,m_l^2,m_l^2),\nonumber\\
D^1_i&=&D_i(0, 2 k_2\cdot p,  0, 2 k_1\cdot p,
 0,m_h^2, m_l^2, m_W^2, m_l^2, m_W^2),\nonumber\\
D^2_i&=&D_i(0, 2 k_1\cdot k_2, 0, 2 k_1\cdot p,
 0,m_h^2, m_l^2, m_W^2, m_W^2, m_W^2),\nonumber\\
D^3_i&=&D_i(0, 0, m_h^2, 0, 2 k_2\cdot p,
 2 k_1\cdot k_2, m_l^2, m_W^2, m_W^2, m_W^2),
\end{eqnarray}
with $C_i$'s and $D_i$'s are the three-point and four-point Feynman integrals defined in Ref. \cite{CiDi}, respectively.

The differential decay rate of $h\to \gamma \nu_l\bar\nu_l$, including both three-point and four-point diagrams contributions, can be expressed as
\begin{eqnarray}\label{decay rate}
\frac{d\Gamma}{dE_\gamma d\cos\theta}&=&\frac{m_h^2E_\gamma^3(m_h-2E_\gamma)}{128\pi^3}
\left[|C_1|^2(1+ \cos^2\theta)+2|C_2|^2\sin^4\left(\theta/2\right)+2|C_3|^2\cos^4\left(\theta/2\right)\right.\nonumber\\
&&\left.+4Re[C_1C_2^*]\sin^4\left(\theta/2\right)+4Re[C_1C_3^*]\cos^4\left(\theta/2\right)\right],
\end{eqnarray}
where $E_\gamma$ is the photon energy in the rest frame of the Higgs boson, and in this frame we have $q^2=m_h^2-2 m_h E_\gamma$. $\theta$ is the angle between the three momentum of the Higgs boson and the three momentum of $\nu$ in the rest frame of the neutrino pair (since the neutrino cannot be observed experimentally, $\theta$ should be integrated out later). The range of these two variables is given by
\begin{eqnarray}
0\leq E_\gamma\leq \frac{m_h}{2},\;\;\;\;-1\leq\cos\theta\leq 1.
\end{eqnarray}
The first term in eq. (\ref{decay rate}) is from the three-point diagrams and will dominant the differential decay rate, the second and third terms are induced from the four-point diagrams while the last two terms are contributed by the interference between three-point and four-point diagrams.

\begin{figure}
  % Requires \usepackage{graphicx}
  \begin{center}
  \includegraphics[width=10cm]{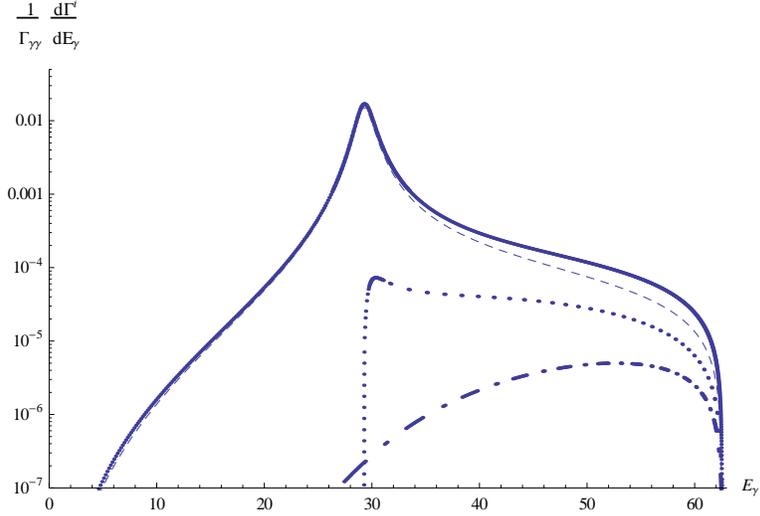}\\
  \caption{The decay spectrum for $h\rightarrow \gamma \nu_e\bar\nu_e$ normalized by $\Gamma(h\rightarrow \gamma\gamma)$(denoted as $\Gamma_{\gamma\gamma}$). The dashed line denotes the contribution of the three-point diagrams, the dotted line shows the behavior of the interference between the three-point and four-point diagrams and the dotted-dashed line the contribution from the four-point diagrams. The real line gives the total contributions.
%The contribution of the interference term is negative when $E_\gamma<29.2 \,\text{GeV}$, in other words, %$\sqrt{q^2}>m_Z$, hence it is not shown in this plot.
}\label{diff}\end{center}
\end{figure}

After integrating over $\cos\theta$ in eq. (\ref{decay rate}), one can get the decay spectrum ${d\Gamma}/{d E_\gamma}$. The $h\rightarrow \gamma \nu_e\bar\nu_e$ decay spectrum, normalized by $\Gamma(h\rightarrow \gamma\gamma)$, has been displayed in Figure \ref{diff}.
Different types of contributions, including the three-point,
four-point diagrams and their interference, are plotted separately for comparison.
From these plots, one can easily find that the three-point diagrams give
the dominant contribution while the contribution of the four-point diagrams
is very small.
It is easy to see that the amplitude (\ref{triangle}) and differential decay rate in eq. (\ref{decay rate}) from the dominant three-point diagrams, if neglecting the small mass of the neutrinos, have no any difference for different types of neutrinos in the final states.
Therefore it is expected that the decay spectrum for the $\nu_\mu$ and $\nu_\tau$ modes will be very similar to the case of the $\nu_e$ mode, as already shown in Figure \ref{diff}. Actually, we have confirmed this point in our numerical analysis.

The decay rate of $h\rightarrow \gamma \nu_l\bar\nu_l$ can be obtained by integrating $E_\gamma$ in the decay spectrum ${d\Gamma}/{d E_\gamma}$, so we get
\begin{eqnarray}
\Gamma(h\to\gamma\nu_e\bar\nu_e)=0.47\,\text{keV},\label{rateemode}
\end{eqnarray}
in which the contribution of the three-point diagrams is $\Gamma(h\to\gamma Z^*\to \gamma\nu_e\bar\nu_e)=0.447\text{keV}$. Since the four-point diagrams give very small contributions, both the charged lepton and neutrino masses can actually be neglected in our numerical calculations. This will lead to the same results as eq. (\ref{rateemode}) for the $\nu_\mu$ and $\nu_\tau$ modes. Experimentally, the neutrinos are invisible, only a single photon and missing energy can be observed,  so the decay rate of the Higgs decays to a photon and invisible particles in the SM is
\begin{eqnarray}
  \Gamma(h\to\gamma+invisible)=\sum_{\nu_l=\nu_e,\nu_\mu,\nu_\tau}
  \Gamma(h\to\gamma\nu_l\bar\nu_l)=1.41\,\text{keV}=15.2\%\,\Gamma
  (h\to\gamma\gamma).\label{rate}
\end{eqnarray}

In conclusion, we have analyzed the rare decay modes $h\rightarrow\gamma \nu_l\bar\nu_l$ with
$\nu_l = \nu_e,\,\nu_\mu$ and $\nu_\tau$ in the SM. It is found that these processes are dominated by the $Z^*$ pole transition $h\to \gamma Z^*\to \gamma \nu_l\bar{\nu}_l$, and four-point box diagrams only give very small contributions. Theoretical prediction of the decay rate in the SM is quite under control, as shown in eqs. (\ref{rateemode}) and (\ref{rate}). Therefore, in the future high statistics experiments, such as Higgs factory, the investigation of the $h\to\gamma+invisible$ decays could be very interesting both to increase our understanding of the properties of SM Higgs boson and to explore the novel Higgs dynamics in new physics scenarios.

\section*{Acknowledgments}
This work was supported in part by the NSF of China under Grant Nos. 11075149 and 11235010.

\end{document}